\documentclass[10pt,twocolumn]{article}
\usepackage[top=1.9cm, bottom=1.9cm, left=1.7cm, right=1.7cm]{geometry}
\usepackage{times}
\usepackage[hyphens]{url}  %
\usepackage[keeplastbox]{flushend}  %
\usepackage{graphicx}  %
\usepackage{xcolor}
\usepackage{flushend}
\usepackage{booktabs}
\usepackage{graphicx}
\usepackage{paralist}
\usepackage{enumitem}

\usepackage[small,bf]{caption}
\usepackage{caption}
\usepackage{subcaption}
\usepackage[hang,flushmargin]{footmisc}

\usepackage[bookmarks=true,%
            bookmarksnumbered=true,%
            colorlinks=true,%
            linkcolor=linkcol,%
            citecolor=citecol,%
            urlcolor=urlcol,%
            hypertexnames=true,%
            pdfpagelabels,
	    ]%
      {hyperref}

\usepackage[OT2,OT1]{fontenc}
\newcommand\cyr
{
\renewcommand\rmdefault{wncyr}
\renewcommand\sfdefault{wncyss}
\renewcommand\encodingdefault{OT2}
\normalfont
\selectfont
}
\DeclareTextFontCommand{\textcyr}{\cyr}

\usepackage[compact]{titlesec}
\titlespacing*{\section}{0pt}{*4}{4pt} 
\titlespacing{\subsection}{0pt}{*3}{3pt}

\definecolor{linkcol}{rgb}{0,0,0.5}
\definecolor{citecol}{rgb}{0,0.5,0.3}
\definecolor{urlcol}{rgb}{0.3,0,0}

\renewenvironment{thebibliography}[1]{
  \begin{oldthebibliography}{#1}
    \setlength{\itemsep}{0.1em}
    \setlength{\parskip}{0.1em}
}
{
  \end{oldthebibliography}
}

\renewcommand{\footnoterule}{%
  \kern -3pt
  \hrule width 1in
  \kern 2pt
}

\usepackage{xspace}
\newcommand{\descr}[1]{\smallskip\noindent\textbf{#1}}

\makeatletter
\def\url@leostyle{%
  \@ifundefined{selectfont}{\def\UrlFont{}}%
  {\def\UrlFont{}}%
}
\makeatother
\usepackage[hyphenbreaks]{breakurl}

\newif\ifwatermark
\watermarkfalse

\ifwatermark
    \usepackage{draftwatermark}
    \SetWatermarkScale{.7}
    \SetWatermarkText{\shortstack{In Submission:\\Do Not Distribute}}
    \SetWatermarkColor[rgb]{1,0.88,0.88}
\fi

\usepackage{etoolbox}
\makeatletter
\patchcmd\@combinedblfloats{\box\@outputbox}{\unvbox\@outputbox}{}{%
   \errmessage{\noexpand\@combinedblfloats could not be patched}%
}%
 \makeatother
\AtBeginShipout{%
  \ifnum\value{page}>1 %
    \typeout{* Additional boxing of page `\thepage'}%
    \setbox\AtBeginShipoutBox=\hbox{\copy\AtBeginShipoutBox}%
  \fi
}

\begin{document}
\title{\LARGE \bf Killing the Password and Preserving Privacy with Device-Centric and Attribute-based Authentication}

\author{Kostantinos Papadamou$^\star$, Savvas Zannettou$^\star$, Bogdan Chifor$^\dagger$, Sorin Teican$^\dagger$, George Gugulea$^\dagger$\\ 
Annamaria Recupero$^\ddagger$, Alberto Caponi$^\ddagger$, Claudio Pisa$^\ddagger$, Giuseppe Bianchi$^\ddagger$, Steven Gevers$^\mp$\\ 
Christos Xenakis$^\pm$, Michael Sirivianos$^\star$ \\[0.5ex]
\normalsize $^\star$Cyprus University of Technology, $^\dagger$certSIGN, $^\ddagger$CNIT - University of Rome Tor Vergata,\\
\normalsize $^\mp$Verizon Enterprise Solutions, $^\pm$University of Piraeus\\
\normalsize \{ck.papadamou, sa.zannettou\}@edu.cut.ac.cy, \{bogdan.chifor, sorin.teican, george.gugulea\}@certsign.ro, \\
\normalsize annamaria.recupero@gmail.com, \{alberto.caponi, claudio.pisa, giuseppe.bianchi\}@uniroma2.it, \\
\normalsize steven.gevers@be.verizon.com, xenakis@unipi.gr, michael.sirivianos@cut.ac.cy
}
\date{}

\maketitle
\begin{abstract}
Current authentication methods on the Web have serious weaknesses. 
First, services heavily rely on the traditional password paradigm, which diminishes the end-users' security and usability. 
Second, the lack of attribute-based authentication does not allow anonymity-preserving access to services. 
Third, users have multiple online accounts that often reflect distinct identity aspects. 
This makes proving combinations of identity attributes hard on the users.

In this paper, we address these weaknesses by proposing a privacy-preserving architecture for device-centric and attribute-based authentication based on: 1) the seamless integration between usable/strong device-centric authentication methods and federated login solutions; 2) the separation of the concerns for Authorization, Authentication, Behavioral Authentication and Identification to facilitate incremental deployability, wide adoption and compliance with NIST assurance levels; and 3) a novel centralized component that allows end-users to perform identity profile and consent management, to prove combinations of fragmented identity aspects, and to perform account recovery in case of device loss. 
To the best of our knowledge, this is the first effort towards fusing the aforementioned techniques under an integrated architecture. 
This architecture effectively deems the password paradigm obsolete with minimal modification on the service provider's software stack.
\end{abstract}

%
%
\section{Introduction}
\label{section_intro}
Authentication on the Web relies on the password paradigm, which was developed during the 60s for accessing monolithic mainframe computers. 
We admit that a 128-bit very complex and long ($\sim$20 characters) password used for a specific service is highly secure when it is only stored in the brain of the user and it is computationally hard to guess. 
However, as the number of Web services increases, the password paradigm entails an inextricable tension between security and usability as users become burdened with memorizing and managing multiple passwords. 
At the same time, passwords can be shoulder-surfed, key-logged, replayed, eavesdropped, brute-forced and phished. 
In addition, password databases can be leaked and even if the service follows security good practices (i.e., hashing and salting the passwords) the attacker can guess the password by performing a dictionary-based brute-force attack.
Over the years, the scientific community repeatedly pinpointed the flaws of the password paradigm~\cite{bonneau2012quest,o2003comparing,adams1999users,florencio2007large}.

\begin{figure*}[t!]
\centering
\includegraphics[width=.8\textwidth]{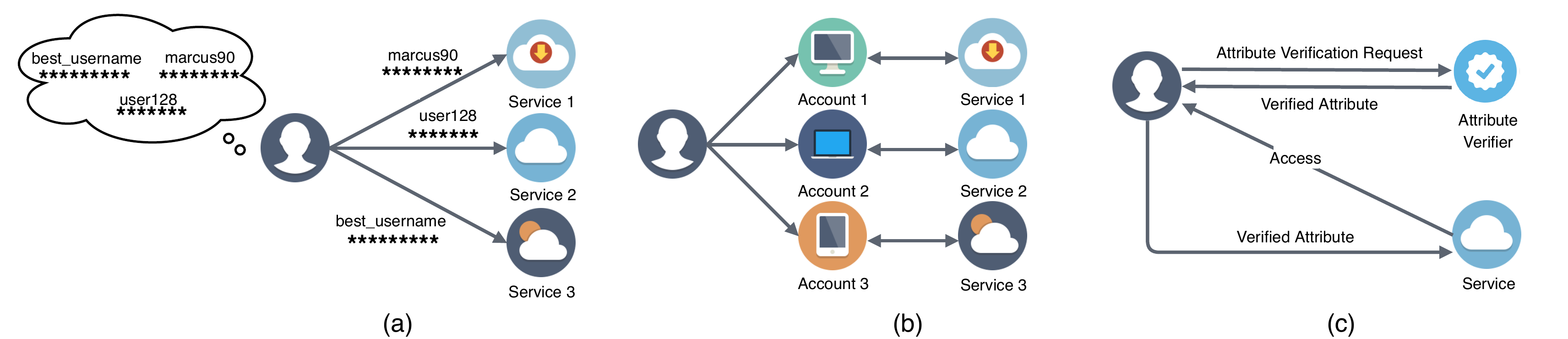}
\caption{Caveats of the prevalent Web authentication paradigm: (a) password overload; (b) identity fragmentation; and (c) lack of support for Attribute-based Access Control (ABAC).}
\label{password_problems} 
\vspace{-1.5em}
\end{figure*}

Fig.~\ref{password_problems} depicts the three main caveats of the currently prevalent Web authentication paradigm. 
First, the password overload problem where users need to remember one secure password for each online service (see Fig. 1(a)). 
As a consequence, they choose easy to remember passwords or resort in re-using the same password across multiple domains~\cite{ives2004domino}.
Second, a user's identity is fragmented across multiple services and there is not an easy way for them to prove account joint-ownership (see Fig. 1(b)). 
Last, there is lack of support for Attribute Based Access Control (ABAC), which facilitates account-less authentication through identity attributes (i.e., age or location); see Fig. 1(c). 
As a result, users are required to reveal multiple aspects of their identity even on services that may only need to verify their age.

Recent efforts aim at mitigating the aforementioned problems by proposing dedicated solutions.
Specifically: 1) federated authentication solutions (i.e., OpenID Connect~\cite{openid2014connect}) alleviate the password overload problem by enabling a Service Provider (SP) to delegate the authentication of end-users to a trusted entity called Identity Provider (IdP); 2) strong and usable password-less authentication mechanisms, such as FIDO UAF~\cite{fido2017specification}; and 3) cryptographic credential stacks that facilitate Privacy-preserving Attribute-based Access Control (PABAC) such as Idemix~\cite{camenisch2002design} and U-Prove~\cite{paguin2013uprove}.
Despite the fact that the aforementioned solutions mitigate the problems to some extent, they suffer from deployability issues as SPs are required to deploy multiple specialized components within their infrastructure.

Other studies~\cite{ross2005stronger, logmeln2019lastpass, siber2019roboform} propose the use of password managers, which enable users to use distinct strong passwords for each online service they use, while the burden of maintaining and remembering the password is offloaded to the password manager. 
However, unlike device-centric authentication with FIDO public-key cryptography, password managers still rely on secret tokens that are susceptible to online guessing, replay, session hijacking, eavesdropping and breach attacks.

In this work, we propose a privacy-preserving federated architecture for device-centric authentication (DCA) that aims to anchor all users' access control needs to devices (i.e., smartphones) that they habitually carry along. 
"Something that end-users almost always have with them," allows users to not have to always "know something for all those accounts they maintain," thus solving the password overload problem.

However, DCA requires special authenticators that most SPs do not have. 
Following recent industry trends, we propose the integration of the design elements proposed by the FIDO Alliance~\cite{fido2017specification} for strong authentication mechanisms, and from the OpenID Foundation for federated authentication~\cite{openid2014connect}. 
This integration enables a federated authentication solution where users are able to authenticate using biometrics (e.g., fingerprint). 
The main advantage of this approach is that the core authentication functionality resides on a trusted entity (IdP), and services (SPs) are able to incrementally adopt this approach with minimal modifications to their infrastructure.

DCA and federation enables the enclosure of strong cryptographic protocols transparent to the user within the device, thus seamlessly supporting anonymity-preserving attribute-based authentication. 
Additionally, the various sensors embedded in mobile devices facilitate behavioral authentication by capturing various behavioral profiles (e.g., gait, keystroke, etc.).
For increased assurance we employ Mobile Connect (MC)~\cite{gsma2019mobileconnect}, which is the equivalent of a secure SIM authenticator where the Mobile Network Operator (MNO) act as an IdP.
Therefore, promoting the device to the main authentication gateway not only eases the user from the burden of remembering multiple complex passwords, but also facilitates technically complex but needed authenticators that make our architecture fully aligned with the latest NIST standards for authentication~\cite{nist2017identityguidelines}.

However, we admit that the mobile device becoming the main authentication gateway is not by itself a universal remedy as it entails serious caveats. 
First, it becomes a single point of failure in case of device loss or failure; we believe that the lack of an efficient device failure/loss recovery mechanism is the main reason passwords are still in use and they have not been replaced by RSA keys. 
Second, the device is vulnerable to hijacking after the user has been authenticated. 
To overcome these issues, we propose a reliable failure recovery framework by leveraging an innovative centralized entity, dubbed Identity Consolidator (IDC), in conjunction with MC authentication and a separate entity for behavioral authentication, called Behavioral Authentication Authority (BAA). 
At the same time, BAA ensures that unauthorized access to services by illegitimate holders of the device is prevented. 
Importantly, besides failure recovery, the IDC also offers real-world and online identity acquisition, identity and privacy management, and allows to prove combinations of fragmented identity aspects, thus solving the identity fragmentation problem.

\descr{Contributions.} 
In summary, we make the following contributions: 
\begin{enumerate}
    \item We demonstrate the merits of the seamless integration between strong/usable password-less authentication methods and federated login solutions under a privacy-preserving architecture.
    \item We offer support for privacy-preserving ABAC on the mobile device.
    \item We propose the separation of concerns for Authentication, Authorization and Behavioral Authentication to IdPs, SPs and BAAs respectively. This enables the incremental deployability of the proposed architecture.
    \item We propose an innovative architectural component, called Identity Consolidator, that solves the identity fragment problem and provides a rich set of features to the user. Specifically, a user can manage the spectrum of his online accounts and define options that will enhance his security, privacy and user experience on the Web.
    \item We propose an innovative failure recovery framework, which is realized through the IDC, and behavioral and MC authentication.
\end{enumerate}

%
%
\section{Terminology}
\label{section_terminology_background}

\descr{User Device.} 
This is the main gateway to get to DCA.
In this work, we assume a user device that is able to utilize recent advances in the field of Trusted Execution Environments (TEE)~\cite{mcgillion2015open}.
This enables the device to securely safeguard cryptographic credentials within its software stack.

\descr{Identity Providers (IdP).} 
IdPs are trusted entities that are responsible for securely maintaining and transferring end-users' identity attributes.
They incorporate strong authentication mechanisms so that they can regulate end-users access. In the context of Privacy-Preserving Attribute-based Access Control, IdPs are responsible for issuing and verifying the end-users' cryptographic credentials.

\descr{Identity Consolidator (IDC).} 
This is a centralized trusted entity that acts as the main IdP and manages all the access control needs of the user.
The user is able to authenticate to the IDC, issue and verify cryptographic credentials, perform failure recovery (in case of lost or damaged device), and lock/unlock its online accounts. 

\descr{Service Providers (SP).} 
These are entities that are responsible only for authorizing end-users to their service.
All other critical operations (i.e., authentication, verification of credentials) are performed by delegating them to trusted entities (IdPs) via Federated solutions, such as OpenID Connect (OIDC). 

\descr{Behavioral Authentication Authorities (BAA).} 
BAAs are special instance of IdPs that offer behavioral authentication to SPs. 
These entities maintain various behavioral profiles for each user that are obtained using signals that are either captured by the user's device or by the BAA itself, depending on the trait type.

%
%
\section{Threat Model and Requirements}
\label{section_requirements}
In this section we define the threat model and the requirements that guide the design and definition of our architecture.

\subsection{Threat Model}
The proposed architecture faces various threats that we must identify. 
We categorize the identified threats according to the main components of our architecture.

\descr{User Device.}
The mobile device of the user is the most vulnerable component in our architecture. 
We admit that the mobile device can be stolen by an attacker who might or might not be able to perform software (i.e., side channel attack) and/or hardware attacks.

\descr{Service and Identity Providers.}
Like every online service, the SPs in our architecture face various threats. 
First, we have to ensure that the access tokens and all the messages exchanged between the server and the clients are protected and will not be disclosed to an attacker during an authentication. 
Second, we assume an attacker who is able to perform Active (Man-in-the-Middle (MitM), Impersonation, Session Hijacking), Cross Site Request Forgery (CSRF), and Replay attacks.
Last, a compromised IdP is another threat.

\descr{User Privacy.} 
User's privacy is of vital importance in our architecture. 
A malicious SP is in the position to infer a user's identity by combining identity attributes revealed in a series of distinct transactions.
Even if standard anonymization practices are performed by the user, if two or more authorized entities (SPs and/or IdPs) are colluding, the user can be identified.

\subsection{Requirements}
To provide a complete solution and address all the aforementioned problems and threats, our architecture should fulfill the following requirements:

\descr{R1: Standards Compliance.}
\label{requirement_r1}
The proposed system should be compliant with open standards. 
This is crucial as it allows incremental deployability, which can lead to the wide adoption of the proposed architecture.

\descr{R2: Ease of deployment.} 
\label{requirement_r2}
SPs participating in our architecture should be able to offer strong authentication mechanisms to their end-users without the need to modify their software stack.

\descr{R3: Identity Federation and Management.} 
\label{requirement_r3}
To combat identity fragmentation, users should have a federated identity on the Web that they can use to prove various attributes of their identity to IdPs and/or SPs and get access to specific resources. 
This requires a centralized entity that will consolidate the various online accounts of a user while enabling him to maintain control over his identity attributes.

\descr{R4: Failure Recovery.}
\label{requirement_r4}
All user access control needs should be anchored to his device, which enables authentication with various usable and cryptographically strong methods. 
The appropriate failure recovery mechanisms should be supported in case of device loss or failure.
This will allow the unobstructed access to online services during unfortunate events.

\descr{R5: Privacy-preserving ABAC.} 
\label{requirement_r5}
In this work, we aim at providing attribute-based authentication while preserving users' privacy. 
In a typical ABAC scenario the SPs should run the appropriate cryptographic verification stacks in order to be able to authenticate specific attributes. 
However, not all SPs are able to run exotic cryptographic stacks. 
Thus, a critical requirement is to enable SPs that do not run cryptographic credentials to support privacy-preserving ABAC. 

\descr{R6: Multi-factor Authentication.} 
\label{requirement_r6}
SPs that provide access to critical resources may require additional authentication from their users for higher assurance. 
Hence, the proposed architecture should offer additional authentication mechanisms for SPs that wish to further verify the identity of a user.

\begin{figure*}[t!]
\centering
\includegraphics[width=0.8\textwidth]{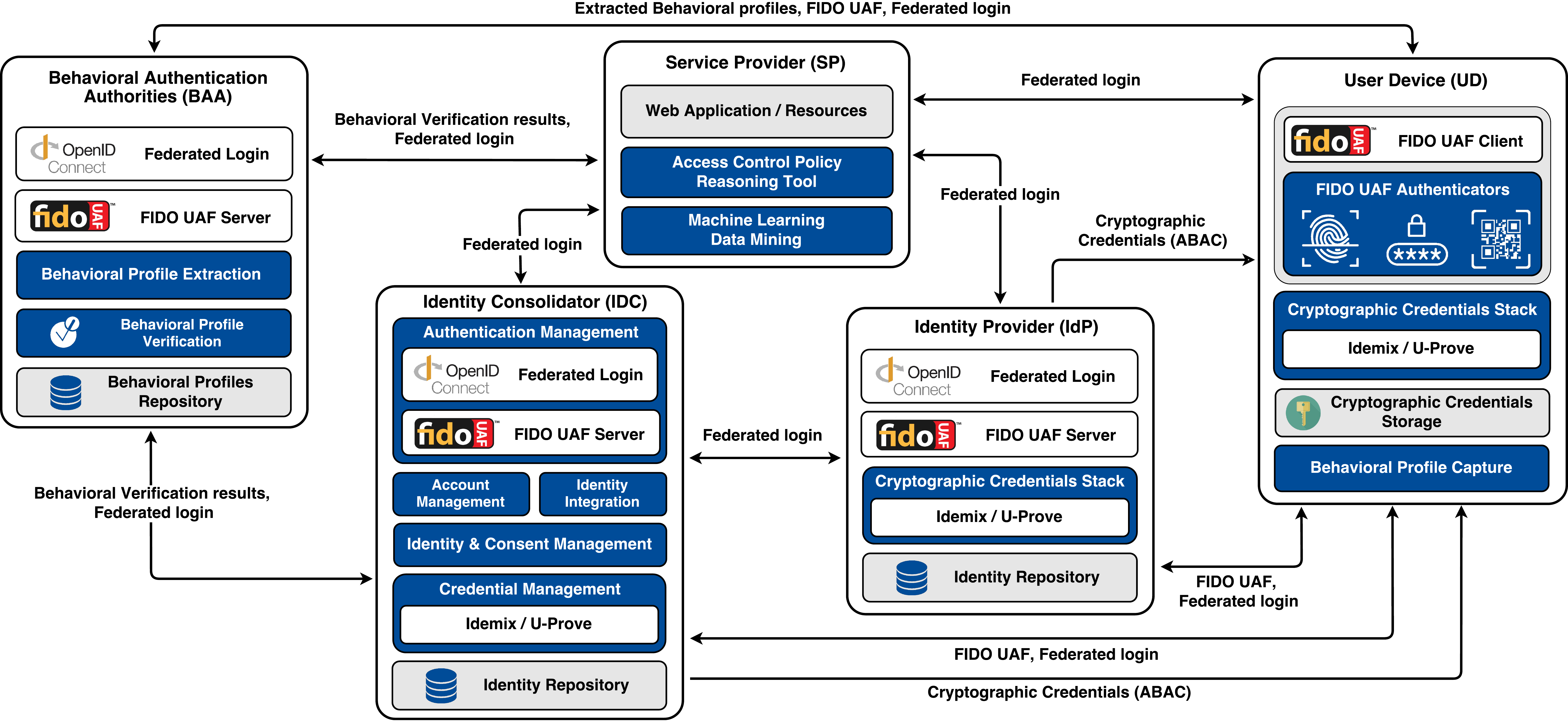}
\caption{Privacy preserving architecture for device-centric and attribute-based authentication. The main architectural components with the modules that comprises each component.}
\label{architecture_diagram} 
\vspace{-1.5em}
\end{figure*}

%
%
\section{Architectural Overview}
\label{section_arch_overview}

In this section we describe the main pillars of our architecture. 
This architecture consists of the following: 1) User Device; 2) Identity Consolidator; 3) Identity Provider; 4) Service Provider; and 5) Behavioral Authentication Authorities. 
Fig.~\ref{architecture_diagram} depicts the proposed architecture including its main components and the interfaces that interconnects them. 
All the communications between the components are built around the OIDC protocol by switching SP and IdP roles. 

\subsection{User Device (UD)}
The mobile device of the user is central in our architecture as we aim to provide DCA. 
We take advantage of the FIDO UAF protocol to make the user's device the main gateway for accessing services on the Web. 
By deploying the FIDO UAF protocol stack we enable human-to-device authentication using biometrics (e.g., fingerprint). 
The device also runs federated authentication protocols (OIDC) with IdPs and SPs (aka, relying parties) for authorization and authentication purposes.

We also deploy cryptographic credential stacks (Idemix and U-Prove) on the device to enable PABAC. 
These stacks allow users to request the issuance of cryptographic credentials from the IDC and/or their IdPs and are responsible for revealing issued credentials to IdPs during an authentication.
The issued credentials are stored in a secure fashion in the Cryptographic Credentials Storage (CCS) that is also part of the device. 
Using a Trusted Execution Environment (TEE) we ensure that credentials stored in the CCS cannot be exported even if the device has been compromised.

Last, to enable continuous and second-factor authentication, the software running on the mobile device includes a module that is responsible for capturing the behavior of the user taking advantage of the various sensors available on the device.

\subsection{Identity Consolidator (IDC)}
The IDC is an integral component in our architecture that fullfils the needs of requirement R3.
It is a centralized fully trusted entity that can be considered as a special instance of an IdP, which offers identity federation, identity and privacy management, and is required for failure recovery. 
The IDC collects identity attributes from various IdPs upon a user's request.
The collected attributes are securely stored in a repository within the IDC. 
We describe below all the modules that comprise the IDC.

\descr{Authentication Management Module (AuthMM).} 
It encapsulates a FIDO-enhanced federated login protocol, which allows the IDC to act as an OIDC IdP for undertaking FIDO authentication. 
This module also allows the IDC to run federated login protocols for transferring identity attributes between distinct IdPs. 
Apart from these, the AuthMM also offers the appropriate failure recovery mechanisms in cases where the user loses access to his device.

\descr{Account Management Module (AMM).} 
The AMM is responsible to keep track of all the BAAs, SPs, and IdPs of a user and it also allows BAA, SP, and IdP admins to register their entities with the IDC. 
Using this knowledge, the AMM acts as a BAA discovery service for SPs that may require a second-factor authentication.
In addition, the AMM enables the user to: 1) manage his IDC account, for example to set his preferred degree of privacy within the IDC (e.g., only certain attribute are stored on the IDC) or completely delete his account; 2) manage the status of his accounts in various SPs and IdPs; and 3) protect his accounts by locking access to them in case of device loss.
The IDC can also act on behalf of the user and lock his online accounts when it detects a high risk of account compromise. 
Last, the AMM facilitates the integration of MC within our architecture. 
To achieve this, IDC act as a relay for SPs that request MC authentication (see Subsection~\ref{subsec:mc_as_service}).

\descr{Credential Management Module (CMM).} 
The CMM enables ABAC in our architecture. 
This module runs cryptographic credential stacks (Idemix/U-Prove) that allows users to issue cryptographic credentials, from their verified identity attributes, directly to their mobile device and then use them to access a variety of SPs. 
The CMM also enables cryptographic credentials management, and allows users to backup their issued credentials at the IDC and restore them anytime on another device in case of device loss or failure.

\descr{Identity Management Module (IMM).} 
IMM consists of the profile and the consent management modules that empower users to manage their identity information. 
The first module provides easy browsing and management of the identity attributes that IdPs and SPs know about a user and informs him about the risks of involuntary attributes inference.
It also allows users to transfer attribute values between different IdPs by extending federated login protocols. 
The latter allows users and IdPs to define consent policies with respect to revealing specific attributes to specified SPs and IdPs.

\descr{Identity Integration module (IIM).} 
\label{subsec:identity_integration_module}
The main responsibility of this module is the standardization and normalization of the users' identity information. 
We acquire this information via physical means (e.g., using Near Field Communication (NFC) to read the user's e-Passport information), and we also perform online identity acquisition where the IDC acts as an SP to receive the users' identity attributes from other IdPs through OIDC.
The IIM encapsulates the required logic for combining, fusing, inferring and validating identity attributes.

\subsection{Identity Providers (IdP)}
Within our architecture, IdPs are entities that authenticate users and share their identity attributes with SPs. 
Each IdP has an identity repository that stores users' attributes.
IdPs also run cryptographic credential stacks (i.e., Idemix and U-Prove) that facilitate the issuance or verification of cryptographic credentials from the stored identity attributes.

\subsection{Service Providers (SP)}
SPs require minimal modifications. 
Namely, they only have to run an OIDC client to communicate with other entities in our architecture. 
SPs are also able to support FIDO and PABAC without the need to run any sophisticated cryptographic stacks by involving IdPs in the authentication process. 
Furthermore, SPs incorporate their business logic within Access Control (AC) policies. 
These policies can be managed by the SP administrator using an Access Control Policy Reasoning tool, which is also responsible to evaluate users' requests on resources based on the defined AC policies.

\subsection{Behavioral Authentication Authorities (BAA)} 
BAAs are separate entities that provide both on-demand and continuous behavioral authentication as part of an entire DCA solution. 
BAAs continuously track the users' behavior through various means and offer a behavioral solution to either SPs or the IDC as a second or third factor authentication. 
Specifically, when requested by an SP or the IDC, BAAs act as an IdP that can verify whether the behavior of a user remains consistent with his usual habits.
The behavioral authentication outcome is released to the aforementioned entities using OIDC.

\subsection{Privacy-Preserving Attribute-based Access Control (PABAC)} 
We enable PABAC by integrating the Idemix~\cite{camenisch2002design} and U-Prove~\cite{paguin2013uprove} cryptographic credential stacks within the OIDC Provider on the IdPs.
In this way, an IdP can act as a credential issuer and/or verifier.
Users can request the issuance of cryptographic credentials by these IdPs or the IDC. 
This solution has various advantages which are: 1) SPs are not required to deploy any cryptographic credential stacks to support PABAC. 
Instead, they delegate the verification of PABAC credentials to IdPs; and 2) it allows for more flexibility as PABAC-enabled IdPs might not be collocated with SPs.

\begin{figure}[t!]
\centering
\includegraphics[width=1.0\columnwidth]{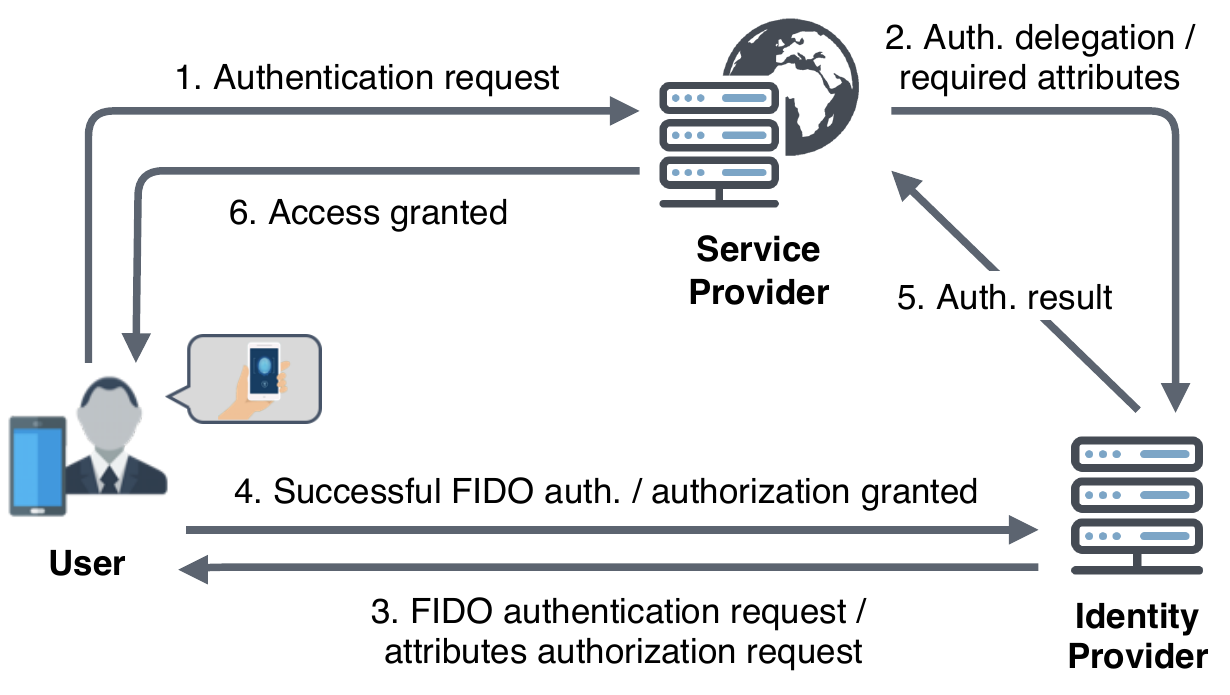}
\caption{FIDO-enhanced federated authentication process.}
\label{fig:fido_oidc_authentication_process} 
\vspace{-1.3em}
\end{figure}

%
%
\section{Design}
\label{section_design}
In this section, we provide adequate information regarding the design of our architecture in which everything is built on top of the OIDC specification. 
We choose to use OIDC with infrastructure authenticator IdPs for incremental deployability.
This is a central design choice that enables us to clearly separate the concerns of SPs and IdPs during an authentication, thus addressing requirements R1 and R2.

\subsection{Diverse Authentication Framework} 
We propose a NIST-compliant~\cite{nist2017identityguidelines} diverse authentication framework, thus addressing requirement R6. 
Specifically, our federated architecture offers various authentication modalities, thus supporting all the assurance levels defined by NIST. 
Depending on which is used, the granted Authenticator Assurance Level (AAL) is determined.
For example, a backup password along with behavioral authentication provides the lower degree of assurance (AAL1), while FIDO authentication alone provides AAL2.
The highest degree of assurance (AAL3) requires a hardware-based cryptographic authenticator and two-factor authentication. 
We achieve this with an enhanced FIDO UAF specification that takes advantage of the TEE that run on end-user devices combined with a secure SIM (Mobile Connect). 
We assume that in the future FIDO and MC will be as secure as a hardware cryptographic token (FIPS 140-2\footnote{\url{https://csrc.nist.gov/publications/detail/fips/140/2/final}}) because of advances in the TEE.

\subsection{FIDO-enhanced Federated Authentication}
OpenID Connect (OIDC) is a simple federated identity layer on top of the OAuth 2.0
protocol~\cite{protocol2019oauth2}, which facilitates federated authentication. 
OIDC enables SPs to delegate the authentication of end-users to IdPs, as well as to obtain profile information about an end-user from the IdPs in an interoperable manner.
The FIDO UAF specification is a password-less solution that enables IdPs to authenticate end-users using strong authenticators (e.g., fingerprint) for user-to-device authentication and cryptographic protocols (e.g., RSA) for device-to-service authentication. 

By combining the concepts of strong authentication alongside with the delegation of authentication to IdPs we allow for a more user-friendly and secure solution for end-users. 
Fig.~\ref{fig:fido_oidc_authentication_process} depicts the proposed FIDO-enhanced federated authentication process.
Initially, when the user tries to authenticate with an SP (step 1), the SP delegates the authentication to an IdP along with a list of identity attributes that the SP requires (step 2).
Then, the IdP requests from the user to authenticate using FIDO on his mobile device (e.g., fingerprint). The IdP also requires from the user authorization to reveal to the SP the requested identity attributes (step 3).
As soon as FIDO authentication is successful and the user has authorized the revelation of the requested attributes (step 4), the IdP informs the SP about the result of the authentication while also providing the requested attribute values (step 5).
At the end, the SP grants to the user access to its resources (step 6).

\begin{figure*}[t!]
\centering
\includegraphics[width=0.9\textwidth]{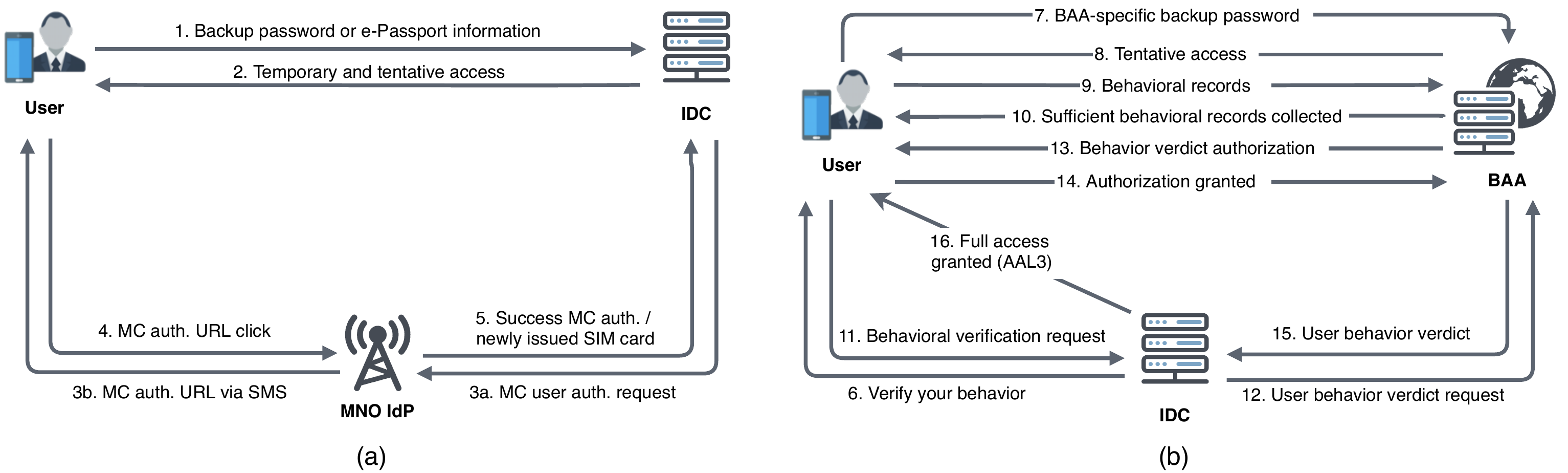}
\caption{
Failure recovery framework: 
(a) shows the first part of the failure recovery that involves MC authentication and (b) shows the second part that involves the verification of the user's behavior through a BAA.
}
\label{fig:failure_recovery_process} 
\vspace{-1.3em}
\end{figure*}

\subsection{Federated Privacy-preserving Attribute-based Authentication}
\label{subsec:federated_pabac}
Our architecture was carefully designed to provide a PABAC solution on top of OIDC, while also addressing requirement R5. 
More precisely, we propose a custom authentication module within the OIDC Provider that acts as an Idemix/U-Prove verifier allowing IdPs to issue and verify cryptographic credentials. 
In fact, we modify the OIDC Provider so that it uses one-time pseudonyms instead of persistent unique identifiers.
In this way, we enable SPs that are not aware of any cryptographic credentials stacks to allow end-users use cryptographic credentials and get access to their resources. 

Federated PABAC offers two concepts of anonymity, namely untraceability and unlinkability.
Untraceability is the security property that precludes the IdP that issued an attribute credential from tracking to which SP the credential has been shown. 
Unlinkability is the property that prevents an IdP or SP from realizing that two or more distinct sessions under the same attribute credential have been initiated by the same user~\cite{camenisch2002design}.
At the same time, users' privacy is preserved since they are able to authenticate to SPs by only revealing the required attributes without revealing their complete identities.

Idemix provides both untraceability and unlinkability, while U-Prove provides only untraceability~\cite{corella2013privacy}.
However, the richer feature set of Idemix comes at a performance cost.
For this reason, we offer both Idemix and U-prove and we allow end-users to choose which one they prefer based on their needs.

\subsection{Mobile Connect (MC) as a Service}
\label{subsec:mc_as_service}
In our architecture we enable SPs to authenticate users using MC. 
Though the IDC we offer MC as a Service, thus allowing incremental deployability of the MC protocol even if the SP is not registered with the MC API Providers. 
To achieve this, the IDC acts as a proxy to SPs for discovering and contacting MC IdPs (MNOs) on behalf of the SP. 
Using OIDC, we see MNOs as any other IdP within our architecture.

The IDC acts as an MC SP to retrieve the required attributes, and then acts as an IdP that proves those attributes to another SP that is not registered in the MC ecosystem. 
In this way, the SPs do not have to be aware of the MC protocol.
They just need to know the value of attributes that can be verified at the required AAL only by MC IdPs. 
Which attributes are those and how they can be retrieved is knowledge that is available only to the IDC.

\subsection{Failure Recovery Framework}
When moving the authentication to the mobile device there are serious caveats that we should consider as we also underline in requirement R4. 
First, in case the device is stolen, the thief has direct access to the secret.
We address this problem via FIDO on devices.
However, the most crucial problem involves recovery after device loss or failure. 
To address this problem, we propose an innovative failure recovery framework, which is realized through the IDC.
More precisely, the IDC federates multiple independent factors (e.g., MC and BAA) that can be easily used in conjunction with a single secure backup password or real-world identity verification to reliably authenticate the user during recovery.

Fig.~\ref{fig:failure_recovery_process} depicts the proposed failure recovery mechanism.
Initially, the user has to authenticate with the IDC using his secure backup password, which is required only for failure recovery. In case the user does not wish to maintain a backup password, he is able to access the IDC with real-world identity verification by scanning his e-Passport using his mobile device (step 1).
If the backup password is correct, or the acquired identity matches the one that he had proved to the IDC before the failure, then he is granted only temporary and tentative access (AAL1) to the IDC, which provides limited functionality (step 2). 
In particular, the user cannot view, restore, or manage PABAC credentials and identity attributes.
Conveniently, with tentative access the user is only able to view his trusted IdPs and initiate authentication to them, the AALs of each IdP, and the backup passwords for the BAAs and his other AAL1 IdPs.

Subsequently, the IDC acts as an SP requesting from the user to authenticate with one of his trusted AAL3 IdPs, for example an MNO IdP via MC (step 3).
Since the user cannot cannot use FIDO to authenticate, he is able to authenticate via SMS\footnote{We acknowledge the vulnerabilities of the SS7-based SMS system~\cite{damien2017ss7sms}.
The authentication to the MNO IdP can also take place in secure ways like FIDO where the public key of the device is installed during the new SIM registration or with a secure version of SMS~\cite{gsma2019mobileconnect}.} using his newly issued SIM card (step 4).
In case of device theft or loss, to ensure that the authentication attempt is performed by the legitimate user, the IDC needs to confirm with the MNO IdP that the given device was reported as lost and a new SIM card was issued (step 5).

Next, for increased assurance, the IDC needs to verify the behavior of the user through one of the trusted BAAs that are registered under his account (step 6).
To do this, the user first authenticates to his BAA using a BAA-specific backup password (step 7).
BAAs can have insecure and easy to memorize backup passwords as their authentication modality is behavioral and the backup password is used only to prevent denial of service attacks. Importantly, in non-tentative mode, the user is able to backup all his BAA-specific and other AAL1 IdPs backup passwords to the IDC which are viewable in tentative mode.
After the user has authenticated, the BAA grants him tentative access and he is not allowed to manage his behavioral profile until his signature is verified as that of the legitimate user's (step 8).
With tentative access, the device sends behavioral records to the BAA, while all the records prior to the new device login are not considered for the authentication (step 9).

Once the BAA has collected sufficient records to give a verdict on whether the user behaves as usual, the user is able to prove his behavior to the IDC (step 10).
When he does so, the IDC acts as an SP while the BAA acts as an IdP authenticating the user based on his behavior and the result is returned to the IDC via OIDC (steps 11-15).
If the verdict is negative the BAA locks that device out of its IdP. 
If the verdict is positive the user is granted full access (AAL3) to the IDC and he is able to issue new FIDO credentials for his account to the new device (step 16).
Both MC and BAA authentication is needed because BAA does not formally increase the NIST authenticator assurance level.

\subsection{De-anonymization Risks and Privacy Assessment}
We extend OIDC so that it keeps a history of the identity attributes revealed to SPs.
Using this information, we provide to the users privacy risk indicators that define the risk of involuntary de-anonymization. 
De-anonymization risk calculation is separated into two categories based on the protocol that a user is using to authenticate: 1) vanilla OIDC; and 2) PABAC.
In the OIDC case, we calculate the probability with which an SP can infer the value of an attribute that the user has not explicitly revealed based on the attributes he has already revealed. 
Due to their nature, Idemix and U-Prove provide unlinkability and untraceability. 
This differentiates the risk calculation from the one performed for OIDC. 
This calculation does not depend on the attributes that the user has shared with an SP in the past since PABAC prevents the SP from linking new sessions with past ones. 
The calculation is made based on the attribute or combination of attributes that the user is about to share with an SP. 
Note that if the user uses untraceable and unlinkable attribute-based authentication, the de-anonymization risk depends on the rarity of the attribute combinations presented to the IdP and SP in a given population and the degree that the SP and IdP know the distribution of attributes in the population.

\subsection{Multi-device Support}
We modify the FIDO UAF client and server software so that it allows the user to register multiple FIDO cryptographic keys, one for each device they use, for each account they maintain. 
This modification enables the users to maintain multiple devices.
Besides this, a user is able to authenticate to an SP through his desktop computer. 
To achieve this, we integrate a Quick Response (QR) authentication server within the IdP's OIDC software to enable authentication from desktop computers to SPs using FIDO. 
Therefore, there is no need for the users to run any user device components on their desktop computers.
We acknowledge that the availability of the mobile device of the user is crucial since a mobile device is required for authentication. 
However, this is also a limitation for FIDO and DCA in general.

\subsection{Deployability and Adoption}
Our federated architecture have many significant benefits for adopters. 
First, user experience is enhanced since a user has to consolidate and prove his identity once at the IDC and then it can be re-used to access multiple IdPs and SPs. 
Second, there is a significant cost reduction to both the end-users (reduction in authenticators) and the SPs (reduction in infrastructure). 
Users do not have to remember dozens of passwords and at the same time they are able to retain their anonymity using PABAC, while SPs can offer FIDO and PABAC authentication to their end-users without the need to deploy any cryptographic credential stacks. 
There is also a significant data minimization for SPs because they do not need to pay for collection and storage of personal identity information. 
As a result, SPs can focus on their mission rather than the business of identity management.

Furthermore, it is clear that IdPs are crucial in federated architectures. 
However, what are the incentives for an organization to play the role of an IdP?
By participating in our architecture, an IdP has many benefits. 
For example, an organization who maintains identity information about users (e.g., age) can offer age verification services to SPs that require age verification from their end-users to abide by the online age verification requirements imposed by regulators (i.e., the Gambling Act 2005\footnote{\url{https://www.legislation.gov.uk/ukpga/2005/19/contents}} for remote gambling in UK).

\begin{figure*}[t!]
\begin{subfigure}[t]{1.0\columnwidth}
    \centering
    \includegraphics[width=1.0\columnwidth]{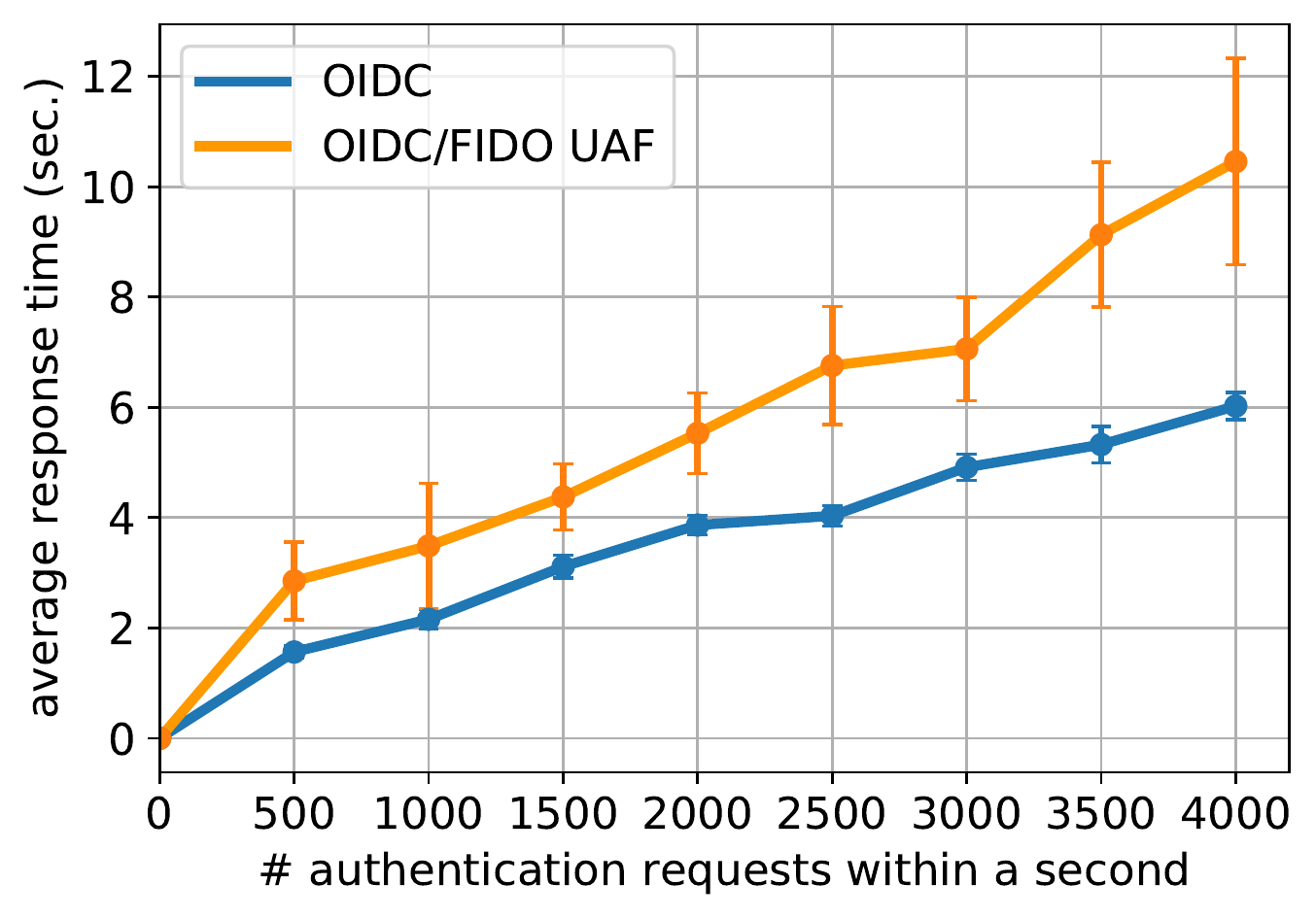}
    \caption{}
    \label{fig:plot_oidc_oidcfido_evaluation}
\end{subfigure}%
~
\begin{subfigure}[t]{1.0\columnwidth}
    \centering
    \includegraphics[width=1.0\columnwidth]{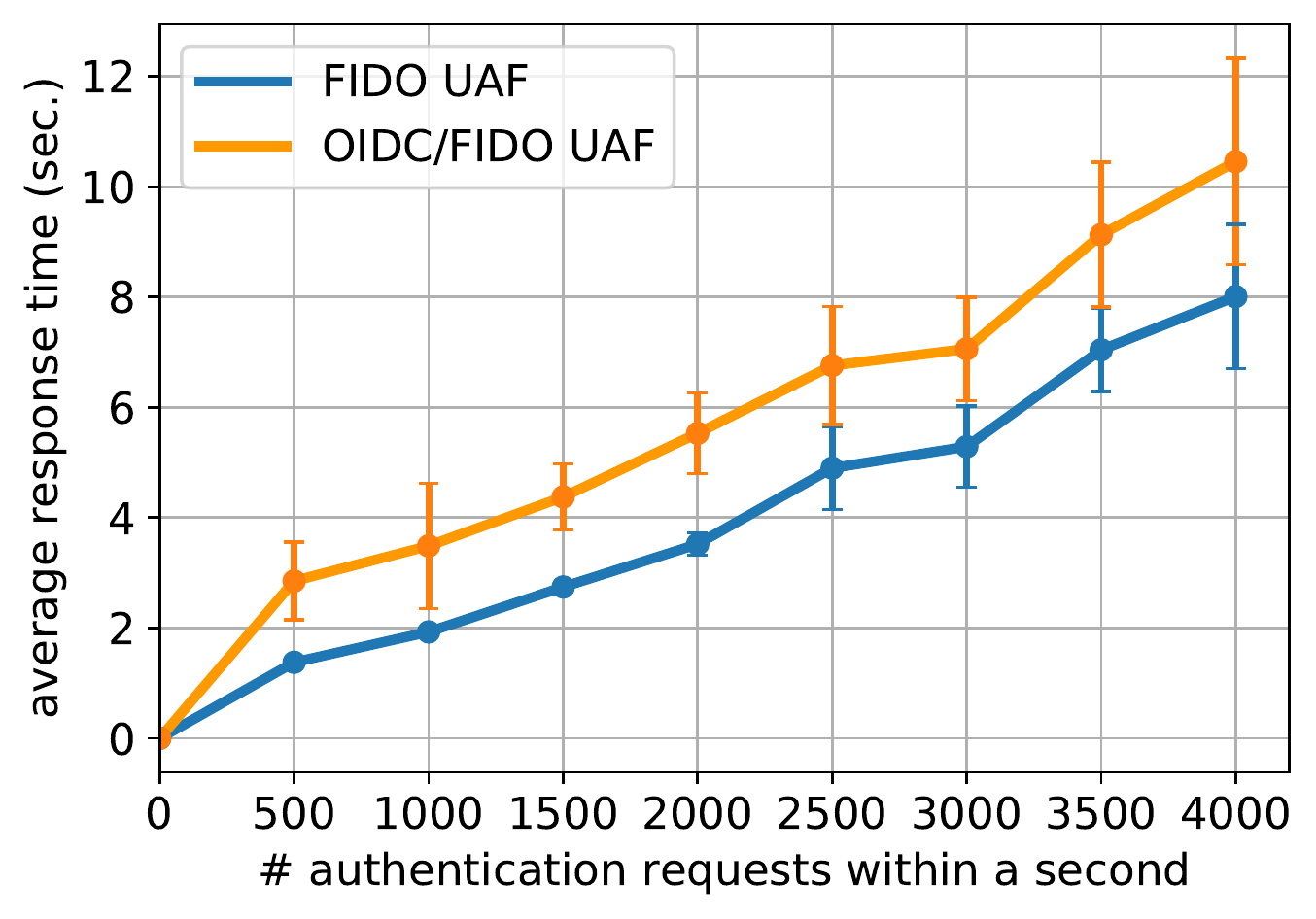}
    \caption{}
    \label{fig:plot_fido_oidcfido_evaluation}
\end{subfigure}%
\caption{
Average response time of an OIDC/FIDO UAF authentication request compared with vanilla: (a) OIDC authentication request; and (b) FIDO UAF authentication request.
}
\label{fig:plots_oidcfido_evaluation} 
\vspace{-1.5em}
\end{figure*}

%
%
\section{Implementation}
\label{section_implementation}
In this section we provide the details of our prototype implementation. 
We implemented all the architecture components as well as all the protocol extensions and integrations that we describe in Section~\ref{section_design}.

\descr{OIDC/FIDO UAF.} 
To exploit the OIDC Provider features, we make use of the OpenAM software\footnote{\url{https://forgerock.org/openam/}}. 
We implemented, within the OIDC Provider, a custom authentication module, which is responsible for undertaking the authentication of the users according to the FIDO UAF specification. 
To achieve this, our custom authentication module communicates with a FIDO UAF Server using a REST interface. 
The FIDO UAF server handles the authentication of the user by communicating with the FIDO UAF client that runs on users' devices.

\descr{OIDC/PABAC.} 
PABAC is realized through the deployment of Idemix/U-Prove credential stacks.
To enable IdPs to act as credentials issuers/verifiers we have implemented a custom authentication module within the OIDC Provider. 
For this purpose we use the FIWARE API\footnote{\url{https://goo.gl/dkG5R8}}, which utilizes both underlying cryptographic protocol stacks used in our architecture. 

\descr{Identity Consolidator.} 
We have implemented the IDC and its respective modules as a Web application. 
Within the IDC we have implemented a well defined REST interface that allows all the other components of our architecture as well as all the external entities to interact with the IDC. 
We have also implemented a custom module within the IDC that allows the IDC to act as an MC proxy. 
This custom module invokes a GSMA API Exchange-enabled~\cite{gsma2019apiexchange} discovery service on a trusted MC Provider. 
This API is mainly used as the federation mechanism for MC authentication.

\descr{User Device.} 
We implemented an Android application that incorporates all the required user device functionality. 
This application runs a FIDO UAF client and utilizes the TEE to store cryptographic (PABAC) credentials. 
We increase maintainability by implementing each module as a separate Android library.

%
%
\section{Evaluation}
\label{section_evaluation}
In this section we evaluate our prototype implementation in terms of performance and security.

%
%
\subsection{Performance Evaluation}
Here we assess the performance of the proposed authentication solution (both OIDC/FIDO and OIDC/PABAC) against the performance of the vanilla OIDC, FIDO UAF, and Idemix protocols.
For a more fair evaluation, when evaluating our federated PABAC authentication solution, we choose Idemix instead of U-Prove because it is the one with the lower performance (see Subsection~\ref{subsec:federated_pabac}).
Each experiment was conducted by sending a batch of authentication requests within a second starting from 500 to 4K requests, while measuring the average response time of the server for each batch of authentication requests, this being the time for all the authentication messages to be exchanged between the client and the server. 
We note that all the authentication requests were successful.

\begin{figure*}[t!]
\begin{subfigure}[t]{1.0\columnwidth}
    \centering
    \includegraphics[width=1.0\columnwidth]{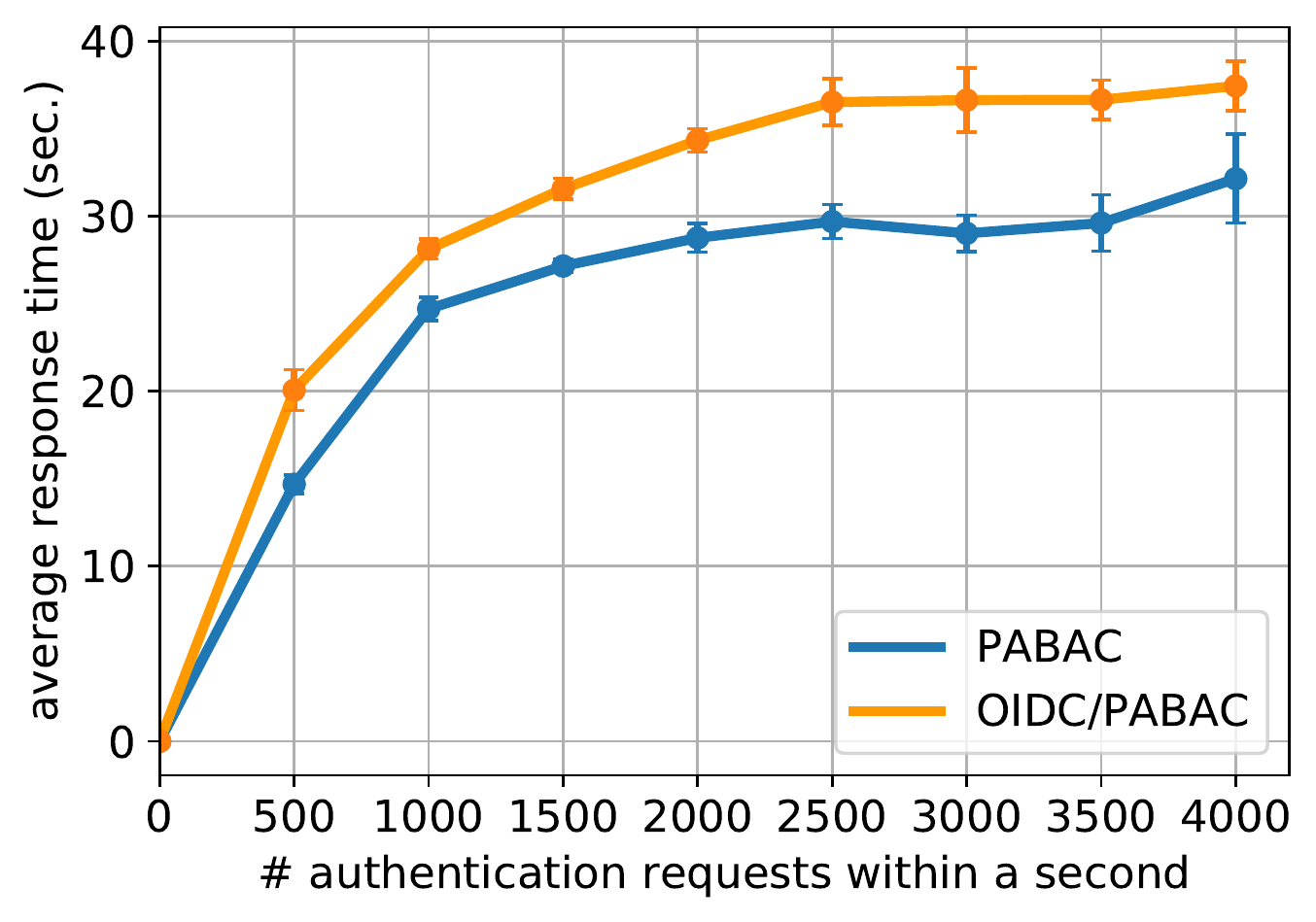}
    \caption{}
    \label{fig:plot_pabac_oidcpabac_evaluation}
\end{subfigure}%
~
\begin{subfigure}[t]{1.0\columnwidth}
    \centering
    \includegraphics[width=1.0\columnwidth]{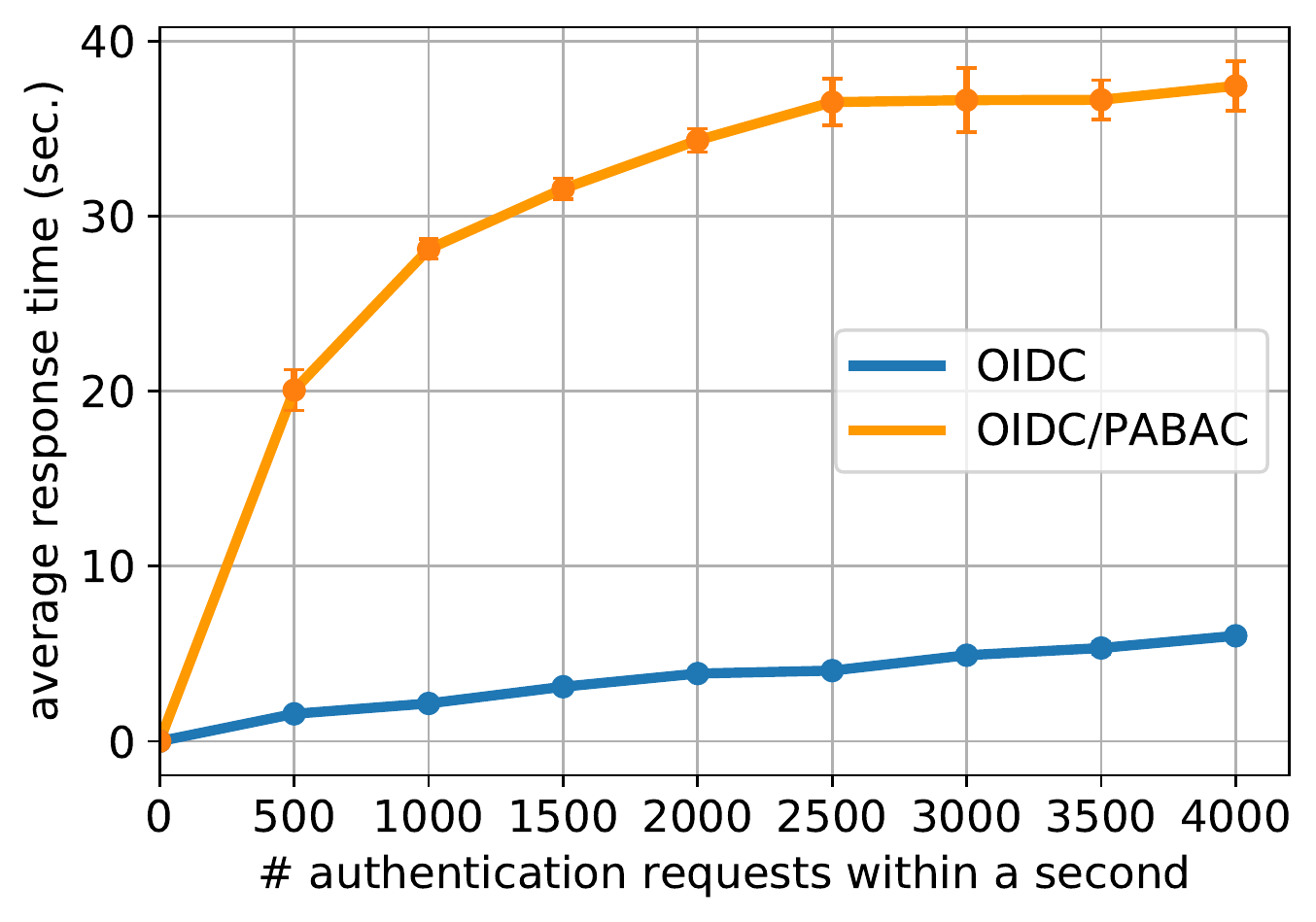}
    \caption{}
    \label{fig:plot_oidc_oidcpabac_evaluation}
\end{subfigure}%
\caption{
Average response time of an OIDC/PABAC authentication request compared with vanilla: (a) Idemix authentication request; and (b) OIDC authentication request.
}
\label{fig:plots_oidcpabac_evaluation} 
\vspace{-1.5em}
\end{figure*}

%
%
\descr{OIDC/FIDO UAF.}
As described in Section \ref{section_implementation}, we implemented a custom authentication module by deploying a FIDO UAF server to the IdP's software stack to enable IdPs authenticate end-users using FIDO. 
Here, we evaluate this deployment in terms of performance to identify: 1) how it performs under high authentication demands; and 
2) the overhead that our custom authentication module introduces compared to the vanilla OIDC and FIDO protocols.
We note that, in our measurements we do not count any user-induced delays (i.e., the time the user needs to enter his password) and we use a 20 characters long password for authentication in the vanilla OIDC case.
We evaluate the performance of a vanilla OIDC deployment by employing the standard OIDC authentication process.
A similar evaluation was also conducted for the vanilla FIDO UAF authentication process. 
All the simulations were performed by porting an Android client on a desktop, which implements the required functionality for standard OIDC and FIDO authentication, and we simulate the parallel authentication processes using different threads. 

Following the same approach, we also evaluate our OIDC/FIDO UAF authentication module. 
Again, the simulations are performed by running our Android OIDC/FIDO client implementation on a desktop. 
We repeat each experiment $10$ times and we calculate the $95\%$ confidence interval of the average response time of each deployment as the number of authentication requests increases.

Figures~\ref{fig:plot_oidc_oidcfido_evaluation} and \ref{fig:plot_fido_oidcfido_evaluation} present the results of the evaluation of our OIDC/FIDO custom authentication module as well as those of vanilla OIDC and FIDO.
We observe that our custom authentication module scales along with the number of authentication requests, with the server's average response time not being drastically impacted.
Compared with the vanilla OIDC and FIDO, our implementation does not introduce any substantial delay in the authentication process: when the number of simultaneous authentication processes is 4K, the average response time of an OIDC/FIDO authentication request is $4.5$ sec and $2.4$ sec more than the average response time of the vanilla OIDC and FIDO protocols, respectively.
Considering the advantages of our proposed solution, we consider the additional delays as negligible.
We note that all the experiments were conducted using an OpenAM software stack and a FIDO UAF server that both run on the IdP and the client requests are executed using an Internet connection.

%
%
\descr{OIDC/PABAC.}
Next, we evaluate our OIDC/PABAC custom authentication module.
For this evaluation we implemented a custom OIDC/PABAC authentication module by utilizing the Idemix credentials stack.
We deployed the implemented module to the IdP's software stack to enable the IdP to act as an Idemix credential issuer and verifier.
The purpose of this evaluation is to identify: 1) how our PABAC-enabled IdP performs under high authentication demands; and 2) the overhead that our implementation introduces compared to the vanilla OIDC and Idemix protocols.
We note that in our measurements we do not count the time required for the issuance of the Idemix credential.

We evaluate both a vanilla Idemix deployment and our OIDC/PABAC implementation.
Similar to the OIDC/FIDO evaluation, we repeat each experiment $10$ times and we calculate the $95\%$ confidence interval of the average response time of each deployment as the number of authentication requests increases.
Figures~\ref{fig:plot_pabac_oidcpabac_evaluation} and~\ref{fig:plot_oidc_oidcpabac_evaluation} present the results of the evaluation of our OIDC/PABAC implementation compared with those of the vanilla Idemix and OIDC protocols, respectively.
We observe that, the average response time of our custom authentication module follows a similar trend to the one of vanilla Idemix authentication and it does not introduce substantial delay in the authentication process: when the number of parallel authentication processes is 4K, the average response time of an OIDC/PABAC authentication request is $5.3$ sec. more than the one of the vanilla Idemix.

%
%
\subsection{Security Analysis}
Here we discuss how we defend against all the possible attacks defined in our threat model.
We categorize the threats and the mitigation strategies that we employ to defend against them according to the main components of our architecture.

\descr{User Device.}
The mobile device of the user is the most critical and vulnerable component in our architecture because it can be stolen. 
First, assuming that we have an attacker who has stolen the device and he is not able to perform software attacks, our architecture is able to effectively defend against such a threat by employing multi-factor authenticators that need to be activated through a biometric (FIDO and continuous behavioral authentication) that can prevent an attacker from being authenticated as the legitimate user; and more importantly by offering a specialized account locking module that is part of the AMM of the IDC allowing the user to lock access to his online accounts on the stolen device.

Next, we also examine the case where the attacker is able to perform software attacks. 
If the behavior capturing protocols run in the Normal World (Rich OS), a skilled software attacker can intervene and modify the contents of the memory while also modify the information captured from the device's sensors. 
However, since all the local measurements are immediately sent to the BAA and are not stored locally on the device then we can prevent such an attacker from bypassing the behavioral authentication. 
On the other hand, if the protocols run in the Secure World (aka Trust-Zone, or Trusted Execution Environment-TEE), no software attacker can compromise the memory and information paths. 
However, we do not have the ability to develop protocols for TEE as the trusted computing base has to be approved by vendors, such as Intel, Samsung, etc. 
We can just invoke specific services of it, such as storing cryptographic keys and performing secure cryptographic operations. 
For example, the activation of the on-demand behavioral authentication with a verifier is triggered through TEE-enabled secure biometrics (e.g., fingerprint). 
This is supplied by FIDO and can protect the user in case the device has been recently stolen and the behavioral signature has yet to change.

Last, when an attacker is able to perform software and hardware attacks then he can bypass the trusted execution and present himself as the legitimate user only if the device is recently stolen, but again the owner of the device can lock access to his online accounts on that device using the AMM.

\descr{Service and Identity Providers.}
SPs and IdPs face various threats.
Initially, by establishing protected sessions between the SPs/IdPs and the users, our solution guarantees that the access tokens are never exposed to unauthorized parties during an authentication.
Next, to defend against Active attacks like MitM, Impersonation, and Session Hijacking attacks we generate access tokens to the authenticated users that are user- and scope- restricted.
To also defend against CSRF attacks we perform header checks to verify the origin of the source and destination for every request while also using CSRF tokens in the communication between the user and the SP.
Also, using TLS for all the communications between the user device and the IdPs/SPs we are able to defend against Replay attacks.
Last, a compromised IdP is not considered since this is a general problem of federated architectures. If an IdP is compromised, it affects the authentication security only of the SPs that relies on that IdP.

\descr{User Privacy.}
Two or more authorized entities (IdPs and SPs) acting maliciously are considered attackers and might be in place to identify a user.
However, we preserve the users' privacy by employing advanced unlinkable and untraceable cryptographic credentials that are used to authenticate with PABAC-enabled IdPs.
Additionally, using the Consent Management module of the IDC, a user has to provide his consent when revealing identity attributes to SPs.
Last, the Profile Management module of the IDC offers to the users privacy risk indicators for each one of their identity attributes. 
These indicators define the risk of involuntary de-anonymization as well as the possibility of an attribute inference.

%
%
\section{Related Work}
\label{section_related_work}
In this section we review existing work on password paradigm alternatives, behavioral authentication, identity federation and management, and attribute-based access control. 

\subsection{Password alternatives}
In the last few years, the research community realized that the password paradigm is not an ideal solution able to cope with user authentication needs on the Web; mainly because of usability and security concerns. 
At the same time, even relatively secure passwords are not replay-resistant authenticators.
Therefore, various studies aim at either replacing the password paradigm or propose solutions that mitigate its caveats.
Specifically, \cite{adams1999users,morris1979password,yan2004password} analyze the usability and security problems of the password paradigm. 
All studies pinpoint the password overload problem which leads users to choose easy to remember passwords or reuse the same password across multiple domains. 

To overcome these issues, password managers like LastPass~\cite{logmeln2019lastpass} and RoboForm~\cite{siber2019roboform} allow users to use a variety of strong passwords for accessing their online services, while the burden of maintaining and remembering them is offloaded to the password manager. 
However, some studies~\cite{chiasson2006usability,zhao2013vulnerability} highlight that the use of password managers introduce new security and usability issues. 
Namely, end-users cannot properly use password managers and this makes them susceptible to various attacks, while the protection mechanisms of several password managers have many security flaws. 
For example, most password managers are protected with a master password. 
If the master password is leaked to an adversary then the password manager becomes a central location for accessing the user's entire online presence. 
In contrast, in our solution a backup password is only required for failure recovery.
Password managers are also susceptible to replay or server breach attacks, while in our solution even if an adversary overhears the challenge-response communication with the IdP, he cannot sign another challenge without the FIDO secure private-key. 
In case of a breach attack the compromised IdP only contains a perfectly useless list of public-keys. 

Other studies propose alternatives to the password paradigm.
Stajano~\cite{stajano2011pico} proposes Pico, a password replacement which relies on hardware tokens. 
At manufacturing time, SPs inject unique keys in each token, which are used for authentication purposes. 
Trusona~\cite{passwordless2015trusona} offers device-centric password-less and multi-factor authentication through a mobile application. 
A user can register by scanning one of his identity documents.

\subsection{Identity Federation and Management}
The past two decades numerous identity federation and management solutions have emerged. 
WSO2 Identity Server~\cite{identityserver2019wso2} is an open source technology that, when integrated within an SP's infrastructure, can offer singe sign-on (SSO), and identity federation and management. 
Unlike WSO2, SPs in our architecture can have the same benefits by just running an OIDC client instead of having to deploy the whole solution into their infrastructure. 
OpenID 2.0~\cite{recordon2006openid} is a user-centric identity management platform in which each account has Identifiers (URI) at one or multiple IdPs, and enables an end-user to prove the possession of such an identifier. 
Users that own the accounts must remember each of their URIs, so some of them are used to access several SPs for validation and authentication of the user. 
If these SPs are malicious, then the users' attributes could be correlated and reveal their identities. 

Other identity management approaches like Liberty Alliance\cite{liberty2019alliance} offer federated user identities in a more privacy-preserving way.
IdPs use pseudonyms or aliases to reference users to the SPs and these pseudonyms are different in each SP. 
One SP cannot directly reference a user in the namespace of another SP, thus preventing malicious SPs from colluding to correlate user identities. 
Inspired by this approach, we extend OIDC to employ pseudonyms so that user anonymity is maintained when they are used in conjunction with privacy-preserving cryptographic credential stacks on the IdP.

Venkatadri et al.~\cite{venkatadri2016strengthening} propose a framework that uses information about identities that is aggregated across multiple domains to reason about their trustworthiness. 
Instead, we deploy more sophisticated algorithms for assessing the trustworthiness of a user's identity with high confidence (see Subsection~\ref{subsec:identity_integration_module}).

\subsection{Behavioral Authentication}
Behavioral authentication provides an extra layer of security above our first factor of authentication. 
Seminal studies have shown that common security authentication mechanisms like PINs or patterns can be enhanced by adding the behavioral factor as another mean of authentication~\cite{zheng2014you,de2012touch}.
Others~\cite{jorgensen2011mouse,feng2012continuous,shi2010implicit,jakobsson2009implicit} continuously track users' behavior for authentication purposes based on various behavior types.
In most of these approaches, the classifier's location is not specified and they do not consider battery, computational, and space limitations.
At the same time, they only tackle observation and impersonation attacks. 
Song et al.~\cite{song2009trustcube} propose TrustCube, a framework that leverages federated authentication schemes to authenticate users based on their behavior on behalf of any SP.
BehavioSec~\cite{biometrics2018behaviosec} offers continuous behavioral authentication software as a service. 
It uses real-time behavioral and statistical analysis tools to resist attacks like account fraud, sharing, and takeover. 
These solutions are typically deployed on the SP and are application-domain-specific.

In our architecture, BAAs are offered as independent entities that can harvest user behavior data from a user's device in a non-intrusive and battery efficient way.
More precisely, we propose an open architecture under which any entity able to capture behavior can offer behavioral authentication via OIDC, while also offering enhanced protection against attackers that manage to compromise the device.

\subsection{Attribute-based Access Control}
Attribute-based access control provides a boolean model in which resources are accessed only if the applicant has the appropriate access attributes as defined by the so-called policies.
This model uses either one of two attribute based encryption (ABE) methods.
Key-policy ABE~\cite{goyal2006attribute} uses the policies to create the applicant keys and uses the attributes to describe the encrypted data. 
Ciphertext-policy ABE~\cite{bethencourt2007ciphertext} uses a tree form access policy, where attributes are the leaves of the tree. 
Ruj et al.~\cite{ruj2012privacy} propose a privacy-preserving access control scheme, in which the attributes of each user belong to multiple key distribution centers~\cite{d2002unconditionally}. 
The user's identity information is stored in the cloud, which acts as the verifier for the users' credentials.
However, user privacy is not protected in the cloud.
Chase~\cite{chase2007multi} introduces a multi-authority KP-ABE scheme that overcomes the drawbacks of a single authority attribute-based system. 
He proposes global identifiers to distinguish different decryptors and allows independent authorities to monitor attributes and secret keys in a distributed storage. 
Later, Chase and Chow~\cite{chase2009improving} propose an improved version of the scheme were a polynomial number of independent authorities is set to monitor attributes and distribute secret keys.

In contrast, in our architecture we integrate cryptographic credentials stacks (Idemix~\cite{camenisch2002design} and U-Prove~\cite{paguin2013uprove}) to let users prove their identity attributes to SPs using cryptographic credentials that are securely stored on their devices. 
In addition, by integrating PABAC with OIDC, we enable any SP to offer PABAC authentication without the need to deploy any cryptographic credential verification stacks.

%
%
\section{Conclusions}
\label{section_conclusions}
In this work we propose an architecture for preserving privacy with device-centric and attribute-based authentication while also solving the serious caveats that the password paradigm has. 
It serves as an alternative for SPs that wish to replace their existing authentication mechanisms without the need to deploy any sophisticated software stacks. 
We readily admit that not all components of our architecture are individually novel. However, combining them together under one architecture, they produce the first proof-of-concept that password-less authentication can be done securely and in a user-friendly fashion under the device-centric paradigm.
Our evaluation results show that our solution can be adopted by end-users and SPs without friction.

%
%
\section*{Acknowledgment}
This research has received funding from the European Union's Horizon 2020 Framework Program under the ReCRED project (GA No. 653417), and the Marie Sk\l{}odowska-Curie INCOGNITO project (GA No. 824015).

\small

\bibliographystyle{abbrv}

\end{document}